%% file: master.tex
\documentclass[aps,superscriptaddress,tightenlines,nofootinbib,floatfix,longbibliography,notitlepage,twocolumn]{revtex4-1}
\usepackage[left=18mm,right=19mm,top=23mm,bottom=16mm]{geometry}
\usepackage{placeins}
\usepackage{bbm}
\usepackage{graphicx}
\usepackage[T1]{fontenc}
\usepackage[utf8]{inputenc}
\usepackage[UKenglish]{babel}
\usepackage{siunitx}
\usepackage[dvipsnames]{xcolor}
\usepackage{listings}
\usepackage{amsmath}
\usepackage{mathtools}
\usepackage{amssymb}
\usepackage{dsfont}
\usepackage{upgreek}
\usepackage{tikz-feynman}
\usepackage{braket}
\usepackage{xfrac}
\usepackage{setspace}
\usepackage[linesnumbered,lined,boxed,commentsnumbered]{algorithm2e}
\usepackage{nameref,zref-xr}
\usepackage{hyperref}
\usepackage{cleveref}
\usepackage{comment}

\zxrsetup{toltxlabel}
\zexternaldocument*{supplementary}

\graphicspath{{figure/},{data/}}

\providecommand{\repositoryInformationSetup}{} % Fallback definition if compiled with `make FINAL=1`.
\repositoryInformationSetup

\input{macros} % input rather than include so we don't create macros.aux
\input{answer}

\input{affiliations}

\begin{document}

\title{The Antiferromagnetic Character of the Quantum Phase Transition in the Hubbard Model on the Honeycomb Lattice}

\author{Johann Ostmeyer}        \affiliation{\bonn}
\author{Evan Berkowitz}         \affiliation{\mcfp}\affiliation{\ias}
\author{Stefan Krieg}           \affiliation{\ias} \affiliation{\jsc} \affiliation{\bonn}
\author{Timo A. L\"{a}hde}      \affiliation{\ias} \affiliation{\ikp}
\author{Thomas Luu}             \affiliation{\ias} \affiliation{\ikp} \affiliation{\bonn}
\author{Carsten Urbach}         \affiliation{\bonn}

\date{\today}

\begin{abstract}
We provide a unified, comprehensive treatment of all operators that contribute to the anti-ferromagnetic, ferromagnetic, and 
charge-density-wave structure factors and order parameters of the hexagonal Hubbard Model. We use the Hybrid Monte Carlo algorithm 
to perform a systematic, carefully controlled analysis in the temporal Trotter error and of the thermodynamic limit. We expect our findings
to improve the consistency of Monte Carlo determinations of critical exponents. We perform a data collapse analysis and determine the critical exponent $\upbeta=\critExp$
for the semimetal-Mott insulator transition in the hexagonal Hubbard Model. Our methods are applicable to a wide range of lattice theories 
of strongly correlated electrons.
\end{abstract}

\maketitle

\allowdisplaybreaks[1]

\input{section/introduction}
\input{section/method}

\input{section/ops}
\input{section/results}

\input{section/discussion}
\input{section/acknowledgements}

\FloatBarrier
\newpage
\onecolumngrid
\appendix
\input{section/stochastic-sources}

\bibliographystyle{apsrev4-1}
\bibliography{cns}

\end{document}

%% file: macros.tex
\newcommand{\repoURL}{https://github.com/evanberkowitz/hubbard-observables}

\newcommand{\Figref}[1]{Figure~\ref{figure:#1}\xspace}

\def\Ref#1{Ref.~\cite{#1}} % Use \def because some LaTeX installations define \Ref, others do not.

\newcommand{\half}{\ensuremath{\frac{1}{2}} }                                   %   1/2
\newcommand{\average}[1]{\ensuremath{\left\langle #1 \right\rangle}\xspace}
\newcommand{\aaverage}[1]{\average{\average{#1}}}
\newcommand{\disconnected}[2]{\ensuremath{\average{#1 #2} - \average{#1}\average{#2}}}

\DeclareMathOperator{\im}{i}
\DeclareMathOperator{\real}{Re}
\DeclareMathOperator{\rt}{Rt}
\DeclareMathOperator{\tr}{Tr}
\DeclareMathOperator{\ti}{Ti}
\DeclareMathOperator{\ord}{\mathcal{O}}

\DeclareMathOperator{\thetaord}{\Theta}
\newcommand{\matr}[2]{\left(\begin{array}{#1}#2\end{array}\right)}

\newcommand{\eto}[1]{\ensuremath{\mathrm{e}^{#1}}}
\newcommand{\ordnung}[1]{\ensuremath{\ord\left(#1\right)}}
\newcommand{\littleo}[1]{\ensuremath{o\left(#1\right)}}
\newcommand{\bigtheta}[1]{\ensuremath{\thetaord\left(#1\right)}}

\newcommand{\erwartung}[1]{\ensuremath{\left\langle#1\right\rangle}}
\newcommand{\pdagger}{{\phantom{\dagger}}}
\newcommand{\kappadelta}{\ensuremath{\tilde{\kappa}}}

\newcommand{\phidelta}{\ensuremath{\tilde{\phi}}}

\newcommand{\transpose}{\ensuremath{{}^{\top}}}

\newcommand{\operator}{\ensuremath{\mathcal{O}}}

\newcommand{\SF}{\ensuremath{S_{F}}\xspace}
\newcommand{\SAF}{\ensuremath{S_{AF}}\xspace}

\newcommand{\qm}{\ensuremath{q_{-}}\xspace}

\let\builtinLaTeX\LaTeX
\def\LaTeX{\builtinLaTeX\xspace}

%% file: answer.tex
   \newcommand{\critUold}{\ensuremath{\num{3.834(14)}}}
  
  \newcommand{\critNuold}{\ensuremath{\num{1.185(43)}}} % z\nu=1/\mu

 \newcommand{\critExpold}{\ensuremath{\num{1.095(37)}}}

   \newcommand{\critU}{\ensuremath{\num{3.835(14)}}}
  
  \newcommand{\critNu}{\ensuremath{\num{1.181(43)}}} % z\nu=1/\mu
\newcommand{\critZeta}{\ensuremath{\num{0.761(5)}}}
 \newcommand{\critExp}{\ensuremath{\num{0.898(37)}}}

%% file: affiliations.tex
\newcommand{\bonn}{
    Helmholtz-Institut f\"{u}r Strahlen- und Kernphysik,
    Rheinische Friedrich-Wilhelms-Universit\"{a}t Bonn, 53115 Bonn, Germany
}

\newcommand{\ikp}{
    Institut f\"{u}r Kernphysik,
    Forschungszentrum J\"{u}lich, 54245 J\"{u}lich, Germany
}

\newcommand{\ias}{
    Institute for Advanced Simulation,
    Forschungszentrum J\"{u}lich, 54245 J\"{u}lich, Germany
}

\newcommand{\jsc}{
    JARA \& J\"{u}lich Supercomputing Center,
    Forschungszentrum J\"{u}lich, 54245 J\"{u}lich, Germany
}

\newcommand{\mcfp}{
    Maryland Center for Fundamental Physics,
    University of Maryland, College Park 20742, USA
}

%% file: section/introduction.tex
%!TEX root =  ../master.tex

The Fermi-Hubbard model---a prototypical model of electrons hopping between lattice sites~\cite{Hubbard:1963}---has a rich phenomenology of strongly-correlated electrons, requiring nonperturbative 
treatment~\cite{White:1989zz}.
On a honeycomb lattice, where it provides a basis for studying electronic properties of carbon nanosystems like graphene and nanotubes, the Hubbard model is expected to exhibit a second-order quantum phase transition between a (weakly coupled) semi-metallic (SM) state and an antiferromagnetic Mott insulating (AFMI) state as a function of the electron-electron coupling~\cite{Paiva2005}.

It is noteworthy that a fully-controlled \emph{ab initio} characterization of the SM-AFMI transition has not yet appeared; Monte Carlo (MC) calculations have not provided unique, 
generally accepted values for the critical exponents~\cite{Otsuka:2015iba,semimetalmott}. Such discrepancies are primarily attributed to the adverse scaling of MC algorithms with spatial system
size $L$ and inverse temperature $\beta$. 
The resulting systematic error is magnified by an incomplete understanding of the extrapolation of operator expectation values to the thermodynamic and temporal continuum limits.

Lattice Monte Carlo (LMC) simulation of Hamiltonian and Lagrangian theories of interacting fermions is a mature  field of study, which is seeing tremendous progress in the areas of novel computer hardware, algorithms, and theoretical developments.
Efficient Hybrid Monte Carlo (HMC) algorithms \cite{Duane:1987de,hasenbusch} designed for theories with dynamical fermions, coupled with GPU-accelerated supercomputing \cite{Clark:2009wm}, now allow for the direct simulation of systems of the same size as those used in realistic condensed-matter experiments and applications~\cite{ulybyshev2021bridging}.

While much of the focus of LMC efforts continues to be on Lattice QCD, the algorithms and methods so developed are now rapidly finding their place among the large variety of Hamiltonian theories in condensed matter physics.
Recently, preliminary studies of nanotubes \cite{Luu:2015gpl,Berkowitz:2017bsn} have appeared and treatments of the Fermion sign problem based on formal developments~\cite{Cristoforetti:2012su,Kanazawa:2014qma} have made promising progress towards first-principles treatments of fullerenes~\cite{leveragingML} and doped systems~\cite{Ulybyshev:2019fte}.
Moreover, our understanding of these algorithms' ergodicity properties~\cite{Wynen:2018ryx} and computational scaling~\cite{Beyl:2017kwp,acceleratingHMC} has also recently been placed on a firm footing.

In this work we seek to remove, using Lattice Monte Carlo (LMC) techniques, the systematic uncertainties which affect determinations of the critical exponents of
the SM-AFMI transition in the honeycomb Hubbard model.
We present a unified, comprehensive, and systematically controlled treatment of the anti-ferromagnetic (AFM), ferromagnetic (FM), and charge-density-wave (CDW) order parameters, 
and confirm the AFM nature of the transition from first principles. Building on our determination of the Mott gap~\cite{semimetalmott} we find that the critical coupling $U_c/\kappa=\critU$ and 
the critical exponents---expected to be the exponents of the SU(2) Gross-Neveu, or chiral Heisenberg, universality class~\cite{gross_neveu_orig,heisenberg_gross_neveu}---to be $\nu=\critNu$ and $\upbeta=\critExp$.

%% file: section/method.tex
%!TEX root =  ../master.tex

\section{Method}
We formulate the grand canonical Hubbard model at half filling in the particle-hole basis and without a bare staggered mass.
Its Hamiltonian reads
\begin{equation}
	H=-\kappa \sum_{\erwartung{x,y}}\left(p^\dagger_{x}p^\pdagger_{y}+h^\dagger_{x}h^\pdagger_{y}\right)+\frac{U}{2}\sum_{x}\rho_x\rho_x\,,\label{eqn_particle_hole_hamiltonian}
\end{equation}
where $p$ and $h$ are fermionic particle and hole annihilation operators, $\kappa$ is the hopping amplitude, $U$ the on-site interaction, and
\begin{equation}
	\rho_x = p^\dagger_x p_x -h^\dagger_x h_x,
	\label{eq:charge_operator}
\end{equation}
is the charge operator.
Using Hasenbusch-accelerated~\cite{hasenbusch,acceleratingHMC} HMC~\cite{Duane:1987de} with the BRS formulation~\cite{Brower:2011av,Brower:2012zd} and a 
mixed time differencing~\cite{Brower:2012zd,Luu:2015gpl,semimetalmott} which has favorable computational scaling and ergodicity properties~\cite{Wynen:2018ryx}, 
we generate ensembles of auxiliary field configurations for different linear spatial extents $L$ with a maximum of $L$=$102$ corresponding to 20,808 lattice sites, interaction strengths $U$ (with fixed hopping $\kappa$), inverse temperatures $\beta = 1/T$, and $N_t$ Trotter steps;\footnote{Throughout this work, we use an upright $\upbeta$ for the critical exponent and a slanted $\beta$ for the inverse temperature.}
See \Ref{semimetalmott} for full details.

We have used these ensembles to compute the Mott gap $\Delta$ as a function of $U$ and $\beta$ to locate a quantum critical point (QCP) at $T = 0$~\cite{semimetalmott} using 
finite-size scaling (FSS)~\cite{newmanb99,dutta_aeppli,Shao_2016}. The single-particle gap $\Delta$ was found to open at $U/\kappa=\critUold$. 
Though this is widely expected to coincide with a semimetal-AFMI transition, we did not characterize the nature of the transition in Ref.~\cite{semimetalmott}.
Instead, we found the correlation length exponent $\nu=\critNuold$ from first principles, and estimated the critical exponent $\upbeta_\text{mf}=\critExpold$ for the staggered magnetization $m_s$ under the AFMI assumption and the \emph{mean-field} expectation $m_s \sim \Delta/U$~\cite{Assaad:2013xua}.
In this work we forgo these assumptions and confirm the AFMI character of the transition, finding $\upbeta=\critExp$ in terms of a FSS analysis in the inverse temperature $\beta$.

Since the two values for $\upbeta$ are strictly incompatible ($5\sigma$ difference), our new results prove that the mean-field approximation is not well suited to extract critical exponents in the given model.

At half-filling, we compute expectation values of one-point and two-point functions of bilinear local operators, 
the spins 
\begin{equation}
S^i_{x} = \half (p^\dagger_x,\, (-1)^x h_x) \sigma^i (p_x,\, (-1)^x h^\dagger_x)\transpose
\end{equation}
and the charge~\eqref{eq:charge_operator}
where the $\sigma^i$ are Pauli matrices and $(-1)^x$ provides a 
minus sign depending on the triangular sublattice of the honeycomb to which the site $x$ belongs; the sign originates in the particle-hole transformation.
For operators, we consider the uniform magnetization 
\begin{equation}
S^i_{+} = \sum_x S^i_{x},
\end{equation}
where spins are summed coherently, and the staggered magnetization 
\begin{equation}
S^i_{-} = \sum_x (-1)^x S^i_{x},
\end{equation}
which computes the difference between the sublattices.
We expect the extensive one-point functions, the ferromagnetic magnetization $\average{S^i_+}$, antiferromagnetic magnetization $\average{S^i_-}$, the total charge $\average{\rho_+}$, and 
the charge separation $\average{\rho_-}$ (and their respective intensive one-point functions) to vanish by symmetry at half-filling for all $\beta$ and $U$.
However, the two-point functions 
\begin{equation}
S^{ii}_{\pm\pm} = \aaverage{S^i_\pm S^i_\pm},
\end{equation}
and
\begin{equation}
Q^2 = \aaverage{\rho_+ \rho_+}, \quad Q_-^2 = \aaverage{\rho_-\rho_-},
\end{equation}
need not vanish at 
half-filling. The double-bracket notation indicates the connected correlator, 
\begin{equation}
\aaverage{\operator_1 \operator_2} = \disconnected{\operator_1}{\operator_2}.
\end{equation}
These quantities scale quadratically with the spatial volume\footnote{Here, $V$ denotes the number of unit cells---half the number of lattice sites. Thus $V=L^2$ in case of an $L\times L$ lattice.} 
$V$ which must be divided out to compute their respective intensive partners, which we denote with lower case letters.  For example, $s^{ii}_{\pm\pm}=S^{ii}_{\pm\pm}/V^2$ and similarly for $q^2$ and $\qm^2$.
A \emph{finite} non-vanishing $s^{ii}_{++}$ ($s^{ii}_{--}$) indicates ferromagnetic (antiferromagnetic) order, while a non-vanishing $\qm^2$ indicates CDW order.\footnote{Susceptibilities, like the AFM susceptibility $\chi_{\text{AF}}$, can be reconstructed from these observables at half-filling.}

\begin{figure*}[ht]
	\input{data/m_s_collapse}
	\input{data/m_s_intensive}
	\caption{Left: Data collapse plot with the optimal parameters of $U_c$, $\nu$, and $\beta$ obtained from a simultaneous collapse fit to the gap $\Delta$ and the order parameter $m_s$.
	Note that the ``outliers'' are due to particularly small $\beta$ and are excluded from the analysis.
	For	$U< U_c \simeq \critU$ the order parameter vanishes.
	Right: The AFMI order parameter (staggered magnetization) $m_s$, with all quantities in units of $\kappa$, after the thermodynamic and continuum limit extrapolations.  
	We also show $m_s(U,\beta = \infty)$ as a solid black line with error band (calculated as in Ref.~\cite{semimetalmott}). The legend from the left plot applies to both.}
	\label{figure:structure factors}
\end{figure*}
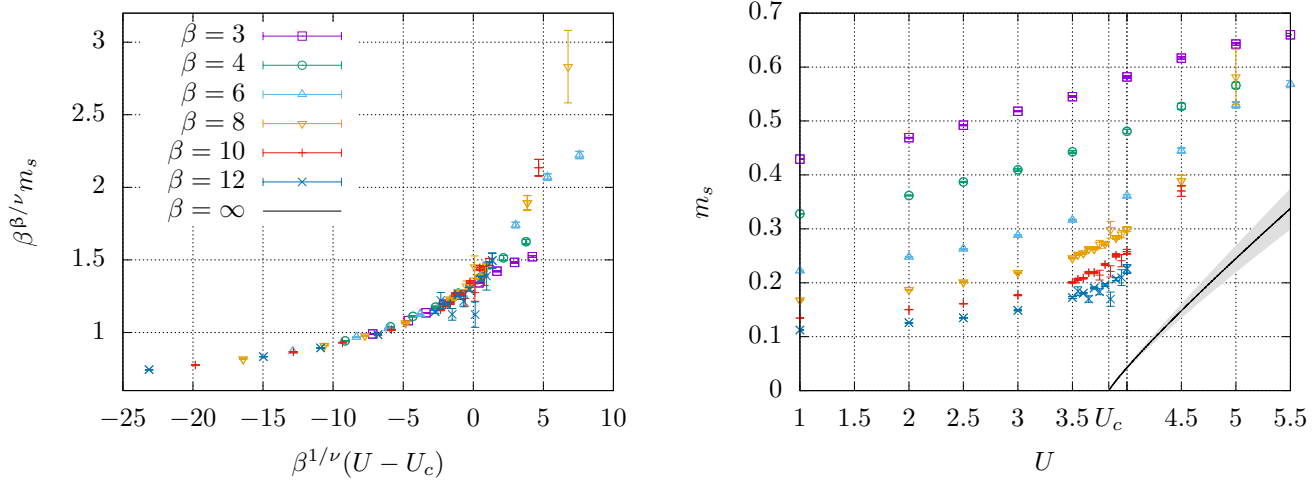

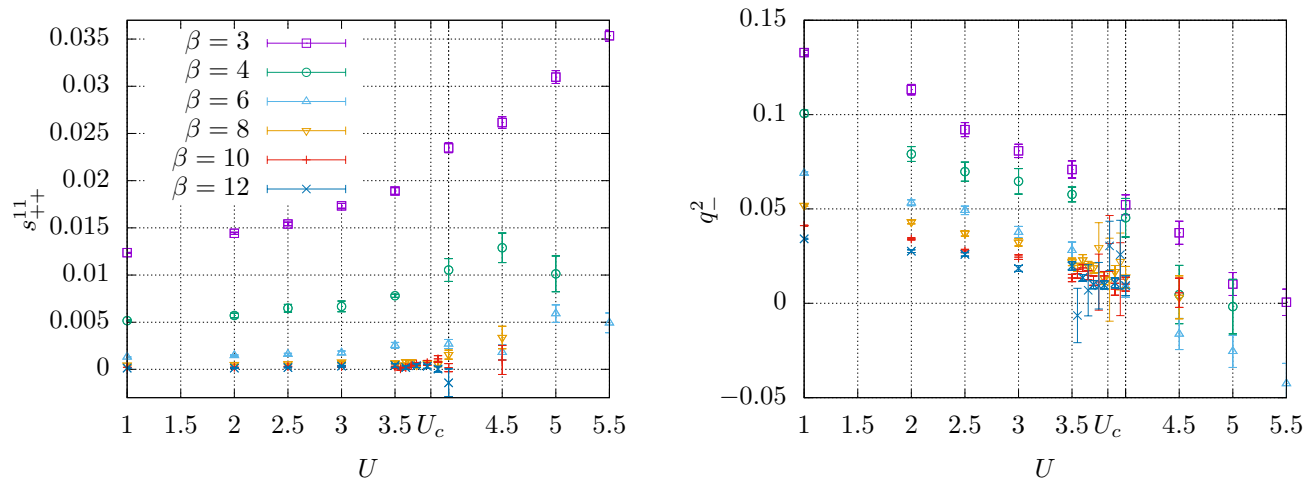
\begin{figure*}[ht]
	\input{data/s_f_1_intensive}
	\input{data/cdw_intensive}
	\caption{Ferromagnetic (left) and charge-density-wave (right) order parameters at different couplings, with all quantities in units of $\kappa$, after the thermodynamic and continuum limit extrapolations. The legend from the left plot applies to both.}
	\label{figure:s_f_and_cdw}
\end{figure*}

On the other hand, at very low temperatures, any finite non-zero value of $s^{ii}_{++}$, $s^{ii}_{--}$, or $\qm^2$ corresponds to an extensive ferromagnetic spin structure factor $\SF=S^{ii}_{++}/V$, 
antiferromagnetic spin structure factor $\SAF=S^{ii}_{--}/V$, or  staggered charge structure factor $S_\mathrm{CDW}=Q_-^2/V$, respectively, diverging linearly with the spatial volume.
The finite temperatures we use, however, provide a natural infrared cutoff for the correlation length, allowing different domains to cancel against one another.
Because the dynamical exponent $z=1$~\cite{PhysRevLett.97.146401} we can obtain intensive order parameters at any finite temperature by taking the thermodynamic limit of these extensive quantities and dividing them by $\beta^2$, rather than the spatial volume $V$. 
Finally, we use finite-size scaling in $\beta$ to remove the infrared regulator and determine zero-temperature properties.

We decompose the observables of interest into a set of operators with zero, one, or two fermion bilinears, each on a definite sublattice, and evaluate each separately.
This decomposition makes it easy to test the operators' individual behaviors in both non-interacting $U\to0$ and non-hopping $\kappa\to 0$ limits, which are analytically known.
We can also verify their expected scaling behaviors in Trotter error and volume.
We reconstruct the operators of physical interest by taking the appropriate linear combination.
%We provide a detailed discussion and list of these operators in \autoref{tab:operators} of~\autoref{sect:15 ops} of the supplemental material.

%% file: data/m_s_collapse.tex
% GNUPLOT: LaTeX picture with Postscript
\begingroup
  % Encoding inside the plot.  In the header of your document, this encoding
  % should to defined, e.g., by using
  % \usepackage[latin1,<other encodings>]{inputenc}
  \inputencoding{latin1}%
  \makeatletter
  \providecommand\color[2][]{%
    \GenericError{(gnuplot) \space\space\space\@spaces}{%
      Package color not loaded in conjunction with
      terminal option `colourtext'%
    }{See the gnuplot documentation for explanation.%
    }{Either use 'blacktext' in gnuplot or load the package
      color.sty in LaTeX.}%
    \renewcommand\color[2][]{}%
  }%
  \providecommand\includegraphics[2][]{%
    \GenericError{(gnuplot) \space\space\space\@spaces}{%
      Package graphicx or graphics not loaded%
    }{See the gnuplot documentation for explanation.%
    }{The gnuplot epslatex terminal needs graphicx.sty or graphics.sty.}%
    \renewcommand\includegraphics[2][]{}%
  }%
  \providecommand\rotatebox[2]{#2}%
  \@ifundefined{ifGPcolor}{%
    \newif\ifGPcolor
    \GPcolortrue
  }{}%
  \@ifundefined{ifGPblacktext}{%
    \newif\ifGPblacktext
    \GPblacktexttrue
  }{}%
  % define a \g@addto@macro without @ in the name:
  \let\gplgaddtomacro\g@addto@macro
  % define empty templates for all commands taking text:
  \gdef\gplbacktext{}%
  \gdef\gplfronttext{}%
  \makeatother
  \ifGPblacktext
    % no textcolor at all
    \def\colorrgb#1{}%
    \def\colorgray#1{}%
  \else
    % gray or color?
    \ifGPcolor
      \def\colorrgb#1{\color[rgb]{#1}}%
      \def\colorgray#1{\color[gray]{#1}}%
      \expandafter\def\csname LTw\endcsname{\color{white}}%
      \expandafter\def\csname LTb\endcsname{\color{black}}%
      \expandafter\def\csname LTa\endcsname{\color{black}}%
      \expandafter\def\csname LT0\endcsname{\color[rgb]{1,0,0}}%
      \expandafter\def\csname LT1\endcsname{\color[rgb]{0,1,0}}%
      \expandafter\def\csname LT2\endcsname{\color[rgb]{0,0,1}}%
      \expandafter\def\csname LT3\endcsname{\color[rgb]{1,0,1}}%
      \expandafter\def\csname LT4\endcsname{\color[rgb]{0,1,1}}%
      \expandafter\def\csname LT5\endcsname{\color[rgb]{1,1,0}}%
      \expandafter\def\csname LT6\endcsname{\color[rgb]{0,0,0}}%
      \expandafter\def\csname LT7\endcsname{\color[rgb]{1,0.3,0}}%
      \expandafter\def\csname LT8\endcsname{\color[rgb]{0.5,0.5,0.5}}%
    \else
      % gray
      \def\colorrgb#1{\color{black}}%
      \def\colorgray#1{\color[gray]{#1}}%
      \expandafter\def\csname LTw\endcsname{\color{white}}%
      \expandafter\def\csname LTb\endcsname{\color{black}}%
      \expandafter\def\csname LTa\endcsname{\color{black}}%
      \expandafter\def\csname LT0\endcsname{\color{black}}%
      \expandafter\def\csname LT1\endcsname{\color{black}}%
      \expandafter\def\csname LT2\endcsname{\color{black}}%
      \expandafter\def\csname LT3\endcsname{\color{black}}%
      \expandafter\def\csname LT4\endcsname{\color{black}}%
      \expandafter\def\csname LT5\endcsname{\color{black}}%
      \expandafter\def\csname LT6\endcsname{\color{black}}%
      \expandafter\def\csname LT7\endcsname{\color{black}}%
      \expandafter\def\csname LT8\endcsname{\color{black}}%
    \fi
  \fi
    \setlength{\unitlength}{0.0500bp}%
    \ifx\gptboxheight\undefined%
      \newlength{\gptboxheight}%
      \newlength{\gptboxwidth}%
      \newsavebox{\gptboxtext}%
    \fi%
    \setlength{\fboxrule}{0.5pt}%
    \setlength{\fboxsep}{1pt}%
\begin{picture}(5040.00,3772.00)%
    \gplgaddtomacro\gplbacktext{%
      \csname LTb\endcsname%%
      \put(814,1142){\makebox(0,0)[r]{\strut{}$1$}}%
      \csname LTb\endcsname%%
      \put(814,1690){\makebox(0,0)[r]{\strut{}$1.5$}}%
      \csname LTb\endcsname%%
      \put(814,2237){\makebox(0,0)[r]{\strut{}$2$}}%
      \csname LTb\endcsname%%
      \put(814,2785){\makebox(0,0)[r]{\strut{}$2.5$}}%
      \csname LTb\endcsname%%
      \put(814,3332){\makebox(0,0)[r]{\strut{}$3$}}%
      \csname LTb\endcsname%%
      \put(946,484){\makebox(0,0){\strut{}$-25$}}%
      \csname LTb\endcsname%%
      \put(1474,484){\makebox(0,0){\strut{}$-20$}}%
      \csname LTb\endcsname%%
      \put(2002,484){\makebox(0,0){\strut{}$-15$}}%
      \csname LTb\endcsname%%
      \put(2530,484){\makebox(0,0){\strut{}$-10$}}%
      \csname LTb\endcsname%%
      \put(3059,484){\makebox(0,0){\strut{}$-5$}}%
      \csname LTb\endcsname%%
      \put(3587,484){\makebox(0,0){\strut{}$0$}}%
      \csname LTb\endcsname%%
      \put(4115,484){\makebox(0,0){\strut{}$5$}}%
      \csname LTb\endcsname%%
      \put(4643,484){\makebox(0,0){\strut{}$10$}}%
    }%
    \gplgaddtomacro\gplfronttext{%
      \csname LTb\endcsname%%
      \put(209,2127){\rotatebox{-270}{\makebox(0,0){\strut{}$\beta^{\upbeta/\nu} m_s$}}}%
      \put(2794,154){\makebox(0,0){\strut{}$\beta^{1/\nu}(U-U_c)$}}%
      \csname LTb\endcsname%%
      \put(1870,3378){\makebox(0,0)[r]{\strut{}$\beta = 3$}}%
      \csname LTb\endcsname%%
      \put(1870,3158){\makebox(0,0)[r]{\strut{}$\beta = 4$}}%
      \csname LTb\endcsname%%
      \put(1870,2938){\makebox(0,0)[r]{\strut{}$\beta = 6$}}%
      \csname LTb\endcsname%%
      \put(1870,2718){\makebox(0,0)[r]{\strut{}$\beta = 8$}}%
      \csname LTb\endcsname%%
      \put(1870,2498){\makebox(0,0)[r]{\strut{}$\beta = 10$}}%
      \csname LTb\endcsname%%
      \put(1870,2278){\makebox(0,0)[r]{\strut{}$\beta = 12$}}%
      \csname LTb\endcsname%%
      \put(1870,2058){\makebox(0,0)[r]{\strut{}$\beta = \infty$}}%
    }%
    \gplbacktext
    \put(0,0){\includegraphics{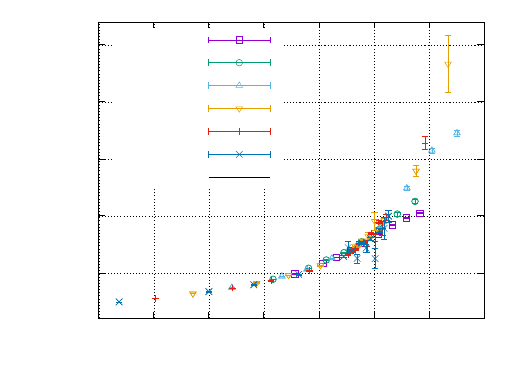}}%
    \gplfronttext
  \end{picture}%
\endgroup

%% file: data/m_s_intensive.tex
% GNUPLOT: LaTeX picture with Postscript
\begingroup
  % Encoding inside the plot.  In the header of your document, this encoding
  % should to defined, e.g., by using
  % \usepackage[latin1,<other encodings>]{inputenc}
  \inputencoding{latin1}%
  \makeatletter
  \providecommand\color[2][]{%
    \GenericError{(gnuplot) \space\space\space\@spaces}{%
      Package color not loaded in conjunction with
      terminal option `colourtext'%
    }{See the gnuplot documentation for explanation.%
    }{Either use 'blacktext' in gnuplot or load the package
      color.sty in LaTeX.}%
    \renewcommand\color[2][]{}%
  }%
  \providecommand\includegraphics[2][]{%
    \GenericError{(gnuplot) \space\space\space\@spaces}{%
      Package graphicx or graphics not loaded%
    }{See the gnuplot documentation for explanation.%
    }{The gnuplot epslatex terminal needs graphicx.sty or graphics.sty.}%
    \renewcommand\includegraphics[2][]{}%
  }%
  \providecommand\rotatebox[2]{#2}%
  \@ifundefined{ifGPcolor}{%
    \newif\ifGPcolor
    \GPcolortrue
  }{}%
  \@ifundefined{ifGPblacktext}{%
    \newif\ifGPblacktext
    \GPblacktexttrue
  }{}%
  % define a \g@addto@macro without @ in the name:
  \let\gplgaddtomacro\g@addto@macro
  % define empty templates for all commands taking text:
  \gdef\gplbacktext{}%
  \gdef\gplfronttext{}%
  \makeatother
  \ifGPblacktext
    % no textcolor at all
    \def\colorrgb#1{}%
    \def\colorgray#1{}%
  \else
    % gray or color?
    \ifGPcolor
      \def\colorrgb#1{\color[rgb]{#1}}%
      \def\colorgray#1{\color[gray]{#1}}%
      \expandafter\def\csname LTw\endcsname{\color{white}}%
      \expandafter\def\csname LTb\endcsname{\color{black}}%
      \expandafter\def\csname LTa\endcsname{\color{black}}%
      \expandafter\def\csname LT0\endcsname{\color[rgb]{1,0,0}}%
      \expandafter\def\csname LT1\endcsname{\color[rgb]{0,1,0}}%
      \expandafter\def\csname LT2\endcsname{\color[rgb]{0,0,1}}%
      \expandafter\def\csname LT3\endcsname{\color[rgb]{1,0,1}}%
      \expandafter\def\csname LT4\endcsname{\color[rgb]{0,1,1}}%
      \expandafter\def\csname LT5\endcsname{\color[rgb]{1,1,0}}%
      \expandafter\def\csname LT6\endcsname{\color[rgb]{0,0,0}}%
      \expandafter\def\csname LT7\endcsname{\color[rgb]{1,0.3,0}}%
      \expandafter\def\csname LT8\endcsname{\color[rgb]{0.5,0.5,0.5}}%
    \else
      % gray
      \def\colorrgb#1{\color{black}}%
      \def\colorgray#1{\color[gray]{#1}}%
      \expandafter\def\csname LTw\endcsname{\color{white}}%
      \expandafter\def\csname LTb\endcsname{\color{black}}%
      \expandafter\def\csname LTa\endcsname{\color{black}}%
      \expandafter\def\csname LT0\endcsname{\color{black}}%
      \expandafter\def\csname LT1\endcsname{\color{black}}%
      \expandafter\def\csname LT2\endcsname{\color{black}}%
      \expandafter\def\csname LT3\endcsname{\color{black}}%
      \expandafter\def\csname LT4\endcsname{\color{black}}%
      \expandafter\def\csname LT5\endcsname{\color{black}}%
      \expandafter\def\csname LT6\endcsname{\color{black}}%
      \expandafter\def\csname LT7\endcsname{\color{black}}%
      \expandafter\def\csname LT8\endcsname{\color{black}}%
    \fi
  \fi
    \setlength{\unitlength}{0.0500bp}%
    \ifx\gptboxheight\undefined%
      \newlength{\gptboxheight}%
      \newlength{\gptboxwidth}%
      \newsavebox{\gptboxtext}%
    \fi%
    \setlength{\fboxrule}{0.5pt}%
    \setlength{\fboxsep}{1pt}%
\begin{picture}(5040.00,3772.00)%
    \gplgaddtomacro\gplbacktext{%
      \csname LTb\endcsname%%
      \put(814,704){\makebox(0,0)[r]{\strut{}$0$}}%
      \csname LTb\endcsname%%
      \put(814,1111){\makebox(0,0)[r]{\strut{}$0.1$}}%
      \csname LTb\endcsname%%
      \put(814,1517){\makebox(0,0)[r]{\strut{}$0.2$}}%
      \csname LTb\endcsname%%
      \put(814,1924){\makebox(0,0)[r]{\strut{}$0.3$}}%
      \csname LTb\endcsname%%
      \put(814,2331){\makebox(0,0)[r]{\strut{}$0.4$}}%
      \csname LTb\endcsname%%
      \put(814,2738){\makebox(0,0)[r]{\strut{}$0.5$}}%
      \csname LTb\endcsname%%
      \put(814,3144){\makebox(0,0)[r]{\strut{}$0.6$}}%
      \csname LTb\endcsname%%
      \put(814,3551){\makebox(0,0)[r]{\strut{}$0.7$}}%
      \csname LTb\endcsname%%
      \put(3274,484){\makebox(0,0){\strut{}$U_c$}}%
      \csname LTb\endcsname%%
      \put(3411,484){\makebox(0,0){\strut{}}}%
      \csname LTb\endcsname%%
      \put(946,484){\makebox(0,0){\strut{}$1$}}%
      \csname LTb\endcsname%%
      \put(1357,484){\makebox(0,0){\strut{}$1.5$}}%
      \csname LTb\endcsname%%
      \put(1768,484){\makebox(0,0){\strut{}$2$}}%
      \csname LTb\endcsname%%
      \put(2178,484){\makebox(0,0){\strut{}$2.5$}}%
      \csname LTb\endcsname%%
      \put(2589,484){\makebox(0,0){\strut{}$3$}}%
      \csname LTb\endcsname%%
      \put(3000,484){\makebox(0,0){\strut{}$3.5$}}%
      \csname LTb\endcsname%%
      \put(3821,484){\makebox(0,0){\strut{}$4.5$}}%
      \csname LTb\endcsname%%
      \put(4232,484){\makebox(0,0){\strut{}$5$}}%
      \csname LTb\endcsname%%
      \put(4643,484){\makebox(0,0){\strut{}$5.5$}}%
    }%
    \gplgaddtomacro\gplfronttext{%
      \csname LTb\endcsname%%
      \put(209,2127){\rotatebox{-270}{\makebox(0,0){\strut{}$m_s$}}}%
      \put(2794,154){\makebox(0,0){\strut{}$U$}}%
    }%
    \gplbacktext
    \put(0,0){\includegraphics{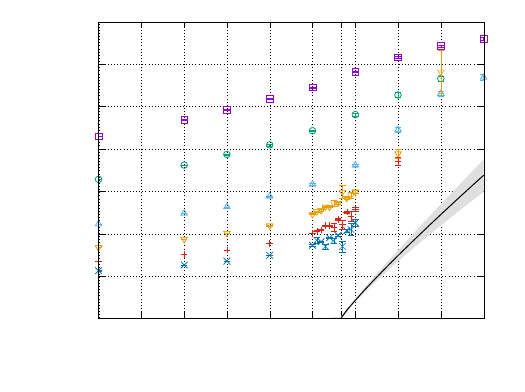}}%
    \gplfronttext
  \end{picture}%
\endgroup

%% file: data/s_f_1_intensive.tex
% GNUPLOT: LaTeX picture with Postscript
\begingroup
  % Encoding inside the plot.  In the header of your document, this encoding
  % should to defined, e.g., by using
  % \usepackage[latin1,<other encodings>]{inputenc}
  \inputencoding{latin1}%
  \makeatletter
  \providecommand\color[2][]{%
    \GenericError{(gnuplot) \space\space\space\@spaces}{%
      Package color not loaded in conjunction with
      terminal option `colourtext'%
    }{See the gnuplot documentation for explanation.%
    }{Either use 'blacktext' in gnuplot or load the package
      color.sty in LaTeX.}%
    \renewcommand\color[2][]{}%
  }%
  \providecommand\includegraphics[2][]{%
    \GenericError{(gnuplot) \space\space\space\@spaces}{%
      Package graphicx or graphics not loaded%
    }{See the gnuplot documentation for explanation.%
    }{The gnuplot epslatex terminal needs graphicx.sty or graphics.sty.}%
    \renewcommand\includegraphics[2][]{}%
  }%
  \providecommand\rotatebox[2]{#2}%
  \@ifundefined{ifGPcolor}{%
    \newif\ifGPcolor
    \GPcolortrue
  }{}%
  \@ifundefined{ifGPblacktext}{%
    \newif\ifGPblacktext
    \GPblacktexttrue
  }{}%
  % define a \g@addto@macro without @ in the name:
  \let\gplgaddtomacro\g@addto@macro
  % define empty templates for all commands taking text:
  \gdef\gplbacktext{}%
  \gdef\gplfronttext{}%
  \makeatother
  \ifGPblacktext
    % no textcolor at all
    \def\colorrgb#1{}%
    \def\colorgray#1{}%
  \else
    % gray or color?
    \ifGPcolor
      \def\colorrgb#1{\color[rgb]{#1}}%
      \def\colorgray#1{\color[gray]{#1}}%
      \expandafter\def\csname LTw\endcsname{\color{white}}%
      \expandafter\def\csname LTb\endcsname{\color{black}}%
      \expandafter\def\csname LTa\endcsname{\color{black}}%
      \expandafter\def\csname LT0\endcsname{\color[rgb]{1,0,0}}%
      \expandafter\def\csname LT1\endcsname{\color[rgb]{0,1,0}}%
      \expandafter\def\csname LT2\endcsname{\color[rgb]{0,0,1}}%
      \expandafter\def\csname LT3\endcsname{\color[rgb]{1,0,1}}%
      \expandafter\def\csname LT4\endcsname{\color[rgb]{0,1,1}}%
      \expandafter\def\csname LT5\endcsname{\color[rgb]{1,1,0}}%
      \expandafter\def\csname LT6\endcsname{\color[rgb]{0,0,0}}%
      \expandafter\def\csname LT7\endcsname{\color[rgb]{1,0.3,0}}%
      \expandafter\def\csname LT8\endcsname{\color[rgb]{0.5,0.5,0.5}}%
    \else
      % gray
      \def\colorrgb#1{\color{black}}%
      \def\colorgray#1{\color[gray]{#1}}%
      \expandafter\def\csname LTw\endcsname{\color{white}}%
      \expandafter\def\csname LTb\endcsname{\color{black}}%
      \expandafter\def\csname LTa\endcsname{\color{black}}%
      \expandafter\def\csname LT0\endcsname{\color{black}}%
      \expandafter\def\csname LT1\endcsname{\color{black}}%
      \expandafter\def\csname LT2\endcsname{\color{black}}%
      \expandafter\def\csname LT3\endcsname{\color{black}}%
      \expandafter\def\csname LT4\endcsname{\color{black}}%
      \expandafter\def\csname LT5\endcsname{\color{black}}%
      \expandafter\def\csname LT6\endcsname{\color{black}}%
      \expandafter\def\csname LT7\endcsname{\color{black}}%
      \expandafter\def\csname LT8\endcsname{\color{black}}%
    \fi
  \fi
    \setlength{\unitlength}{0.0500bp}%
    \ifx\gptboxheight\undefined%
      \newlength{\gptboxheight}%
      \newlength{\gptboxwidth}%
      \newsavebox{\gptboxtext}%
    \fi%
    \setlength{\fboxrule}{0.5pt}%
    \setlength{\fboxsep}{1pt}%
\begin{picture}(5040.00,3772.00)%
    \gplgaddtomacro\gplbacktext{%
      \csname LTb\endcsname%%
      \put(876,918){\makebox(0,0)[r]{\strut{}$0$}}%
      \csname LTb\endcsname%%
      \put(876,1273){\makebox(0,0)[r]{\strut{}$0.005$}}%
      \csname LTb\endcsname%%
      \put(876,1629){\makebox(0,0)[r]{\strut{}$0.01$}}%
      \csname LTb\endcsname%%
      \put(876,1985){\makebox(0,0)[r]{\strut{}$0.015$}}%
      \csname LTb\endcsname%%
      \put(876,2341){\makebox(0,0)[r]{\strut{}$0.02$}}%
      \csname LTb\endcsname%%
      \put(876,2697){\makebox(0,0)[r]{\strut{}$0.025$}}%
      \csname LTb\endcsname%%
      \put(876,3053){\makebox(0,0)[r]{\strut{}$0.03$}}%
      \csname LTb\endcsname%%
      \put(876,3409){\makebox(0,0)[r]{\strut{}$0.035$}}%
      \csname LTb\endcsname%%
      \put(3297,484){\makebox(0,0){\strut{}$U_c$}}%
      \csname LTb\endcsname%%
      \put(3431,484){\makebox(0,0){\strut{}}}%
      \csname LTb\endcsname%%
      \put(1008,484){\makebox(0,0){\strut{}$1$}}%
      \csname LTb\endcsname%%
      \put(1412,484){\makebox(0,0){\strut{}$1.5$}}%
      \csname LTb\endcsname%%
      \put(1816,484){\makebox(0,0){\strut{}$2$}}%
      \csname LTb\endcsname%%
      \put(2220,484){\makebox(0,0){\strut{}$2.5$}}%
      \csname LTb\endcsname%%
      \put(2624,484){\makebox(0,0){\strut{}$3$}}%
      \csname LTb\endcsname%%
      \put(3027,484){\makebox(0,0){\strut{}$3.5$}}%
      \csname LTb\endcsname%%
      \put(3835,484){\makebox(0,0){\strut{}$4.5$}}%
      \csname LTb\endcsname%%
      \put(4239,484){\makebox(0,0){\strut{}$5$}}%
      \csname LTb\endcsname%%
      \put(4643,484){\makebox(0,0){\strut{}$5.5$}}%
    }%
    \gplgaddtomacro\gplfronttext{%
      \csname LTb\endcsname%%
      \put(271,2127){\rotatebox{-270}{\makebox(0,0){\strut{}$s^{11}_{++}$}}}%
      \put(2825,154){\makebox(0,0){\strut{}$U$}}%
      \csname LTb\endcsname%%
      \put(1932,3378){\makebox(0,0)[r]{\strut{}$\beta = 3$}}%
      \csname LTb\endcsname%%
      \put(1932,3158){\makebox(0,0)[r]{\strut{}$\beta = 4$}}%
      \csname LTb\endcsname%%
      \put(1932,2938){\makebox(0,0)[r]{\strut{}$\beta = 6$}}%
      \csname LTb\endcsname%%
      \put(1932,2718){\makebox(0,0)[r]{\strut{}$\beta = 8$}}%
      \csname LTb\endcsname%%
      \put(1932,2498){\makebox(0,0)[r]{\strut{}$\beta = 10$}}%
      \csname LTb\endcsname%%
      \put(1932,2278){\makebox(0,0)[r]{\strut{}$\beta = 12$}}%
    }%
    \gplbacktext
    \put(0,0){\includegraphics{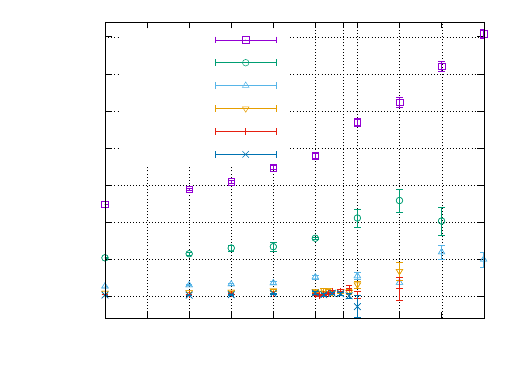}}%
    \gplfronttext
  \end{picture}%
\endgroup

%% file: data/cdw_intensive.tex
% GNUPLOT: LaTeX picture with Postscript
\begingroup
  % Encoding inside the plot.  In the header of your document, this encoding
  % should to defined, e.g., by using
  % \usepackage[latin1,<other encodings>]{inputenc}
  \inputencoding{latin1}%
  \makeatletter
  \providecommand\color[2][]{%
    \GenericError{(gnuplot) \space\space\space\@spaces}{%
      Package color not loaded in conjunction with
      terminal option `colourtext'%
    }{See the gnuplot documentation for explanation.%
    }{Either use 'blacktext' in gnuplot or load the package
      color.sty in LaTeX.}%
    \renewcommand\color[2][]{}%
  }%
  \providecommand\includegraphics[2][]{%
    \GenericError{(gnuplot) \space\space\space\@spaces}{%
      Package graphicx or graphics not loaded%
    }{See the gnuplot documentation for explanation.%
    }{The gnuplot epslatex terminal needs graphicx.sty or graphics.sty.}%
    \renewcommand\includegraphics[2][]{}%
  }%
  \providecommand\rotatebox[2]{#2}%
  \@ifundefined{ifGPcolor}{%
    \newif\ifGPcolor
    \GPcolortrue
  }{}%
  \@ifundefined{ifGPblacktext}{%
    \newif\ifGPblacktext
    \GPblacktexttrue
  }{}%
  % define a \g@addto@macro without @ in the name:
  \let\gplgaddtomacro\g@addto@macro
  % define empty templates for all commands taking text:
  \gdef\gplbacktext{}%
  \gdef\gplfronttext{}%
  \makeatother
  \ifGPblacktext
    % no textcolor at all
    \def\colorrgb#1{}%
    \def\colorgray#1{}%
  \else
    % gray or color?
    \ifGPcolor
      \def\colorrgb#1{\color[rgb]{#1}}%
      \def\colorgray#1{\color[gray]{#1}}%
      \expandafter\def\csname LTw\endcsname{\color{white}}%
      \expandafter\def\csname LTb\endcsname{\color{black}}%
      \expandafter\def\csname LTa\endcsname{\color{black}}%
      \expandafter\def\csname LT0\endcsname{\color[rgb]{1,0,0}}%
      \expandafter\def\csname LT1\endcsname{\color[rgb]{0,1,0}}%
      \expandafter\def\csname LT2\endcsname{\color[rgb]{0,0,1}}%
      \expandafter\def\csname LT3\endcsname{\color[rgb]{1,0,1}}%
      \expandafter\def\csname LT4\endcsname{\color[rgb]{0,1,1}}%
      \expandafter\def\csname LT5\endcsname{\color[rgb]{1,1,0}}%
      \expandafter\def\csname LT6\endcsname{\color[rgb]{0,0,0}}%
      \expandafter\def\csname LT7\endcsname{\color[rgb]{1,0.3,0}}%
      \expandafter\def\csname LT8\endcsname{\color[rgb]{0.5,0.5,0.5}}%
    \else
      % gray
      \def\colorrgb#1{\color{black}}%
      \def\colorgray#1{\color[gray]{#1}}%
      \expandafter\def\csname LTw\endcsname{\color{white}}%
      \expandafter\def\csname LTb\endcsname{\color{black}}%
      \expandafter\def\csname LTa\endcsname{\color{black}}%
      \expandafter\def\csname LT0\endcsname{\color{black}}%
      \expandafter\def\csname LT1\endcsname{\color{black}}%
      \expandafter\def\csname LT2\endcsname{\color{black}}%
      \expandafter\def\csname LT3\endcsname{\color{black}}%
      \expandafter\def\csname LT4\endcsname{\color{black}}%
      \expandafter\def\csname LT5\endcsname{\color{black}}%
      \expandafter\def\csname LT6\endcsname{\color{black}}%
      \expandafter\def\csname LT7\endcsname{\color{black}}%
      \expandafter\def\csname LT8\endcsname{\color{black}}%
    \fi
  \fi
    \setlength{\unitlength}{0.0500bp}%
    \ifx\gptboxheight\undefined%
      \newlength{\gptboxheight}%
      \newlength{\gptboxwidth}%
      \newsavebox{\gptboxtext}%
    \fi%
    \setlength{\fboxrule}{0.5pt}%
    \setlength{\fboxsep}{1pt}%
\begin{picture}(5040.00,3772.00)%
    \gplgaddtomacro\gplbacktext{%
      \csname LTb\endcsname%%
      \put(876,704){\makebox(0,0)[r]{\strut{}$-0.05$}}%
      \csname LTb\endcsname%%
      \put(876,1416){\makebox(0,0)[r]{\strut{}$0$}}%
      \csname LTb\endcsname%%
      \put(876,2128){\makebox(0,0)[r]{\strut{}$0.05$}}%
      \csname LTb\endcsname%%
      \put(876,2839){\makebox(0,0)[r]{\strut{}$0.1$}}%
      \csname LTb\endcsname%%
      \put(876,3551){\makebox(0,0)[r]{\strut{}$0.15$}}%
      \csname LTb\endcsname%%
      \put(3297,484){\makebox(0,0){\strut{}$U_c$}}%
      \csname LTb\endcsname%%
      \put(3431,484){\makebox(0,0){\strut{}}}%
      \csname LTb\endcsname%%
      \put(1008,484){\makebox(0,0){\strut{}$1$}}%
      \csname LTb\endcsname%%
      \put(1412,484){\makebox(0,0){\strut{}$1.5$}}%
      \csname LTb\endcsname%%
      \put(1816,484){\makebox(0,0){\strut{}$2$}}%
      \csname LTb\endcsname%%
      \put(2220,484){\makebox(0,0){\strut{}$2.5$}}%
      \csname LTb\endcsname%%
      \put(2624,484){\makebox(0,0){\strut{}$3$}}%
      \csname LTb\endcsname%%
      \put(3027,484){\makebox(0,0){\strut{}$3.5$}}%
      \csname LTb\endcsname%%
      \put(3835,484){\makebox(0,0){\strut{}$4.5$}}%
      \csname LTb\endcsname%%
      \put(4239,484){\makebox(0,0){\strut{}$5$}}%
      \csname LTb\endcsname%%
      \put(4643,484){\makebox(0,0){\strut{}$5.5$}}%
    }%
    \gplgaddtomacro\gplfronttext{%
      \csname LTb\endcsname%%
      \put(271,2127){\rotatebox{-270}{\makebox(0,0){\strut{}$q^2_{-}$}}}%
      \put(2825,154){\makebox(0,0){\strut{}$U$}}%
    }%
    \gplbacktext
    \put(0,0){\includegraphics{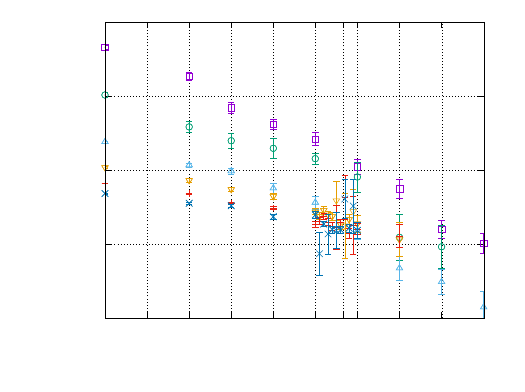}}%
    \gplfronttext
  \end{picture}%
\endgroup

%% file: section/ops.tex
%!TEX root =  ../master.tex
\subsection{Constructing spin structure factors from fundamental operators\label{sect:15 ops}}

For a fully systematic treatment of the different spin operators, we express these in terms of 15 operators $O_{0\dots14}$ of one or two bilinear operators.
Our basis of operators is complete for translationally-invariant equal-time observables (the generalization is straightforward) and we can express the observables of physical interest as linear combinations of the operators in our basis.

Our basis fixes one position on a sublattice.
We choose the operators with odd (even) index to act on $x\in A\ (B)$.
As we do not break the sublattice symmetry explicitly, we can always consider these pairs of operators simultaneously.
An overview of the $O_i$ based on the fermion matrix from~\cite{semimetalmott}
\begin{equation}
	\begin{split}
		M^{AA}_{(x,t)(y,t')} & = \delta_{xy}^{} \left(\delta_{t+1,t^\prime}^{} - \delta_{t,t^\prime}^{} \exp(-i\phidelta_{x,t}^{}) \right), \\
		M^{BB}_{(x,t)(y,t')} & = \delta_{xy}^{} \left(\delta_{t,t^\prime} - \delta_{t-1,t'} \exp(-i\phidelta_{x,t}^{}) \right), \\
		M^{AB}_{(x,t)(y,t')} = M^{BA}_{(x,t)(y,t^\prime)} & =- \kappadelta\,\delta_{\erwartung{x,y}} \delta_{t,t^\prime},
	\end{split}
\end{equation}
is given in Tab.~\ref{tab:operators}, and the various order parameters and structure factors are given in terms of the $O_i$ in Tab.~\ref{tab:observables}, along with the exact values of the physical observables in the $\kappa = 0$ and $U = 0$ limits and their leading error terms.
It should be noted that maximal ordering is not reached for $\beta < \infty$, thus the observables for $\kappa = 0$ are  
limited by the temporal correlation length (the largest inverse Matsubara frequency $\beta/\pi$).

\begin{table*}[ht]
\begin{center}
\caption{Extensive operators $O_i$. Spin comes with a factor $1/2$, therefore the linear operators $O_{1,2}$ have to be divided by 2 and the quadratic ones $O_{3,\dots,14}$ by 4. 
Sums over repeated indices are implicit, so is the addition of the term with $p\leftrightarrow h$.
The two sites $x$ and $y$ are on the same lattice if $x\sim y$.
We define $\ti_x M\coloneqq\tr_S M^{-1}$ and $\rt_x M\coloneqq \real\tr_S M^{-1}$ where $S$ is the sublattice containing $x$. Analogously $\mathds{P}_x$ 
is the projector onto the sublattice containing $x$. The error estimation comes from an exact calculation of $\rt_xM$ at finite temperature and discretization 
in the non-interacting limit.
The exact values in the two limits $U=0$ and $\kappa=0$ are given for zero temperature.
We use the Bachmann–Landau notation to denote that an error in $\bigtheta x$ scales exactly as $x$, an error in $\ordnung{x}$ scales maximally as $x$, and an error in $\littleo{x}$ is asymptotically smaller than $x$.
\label{tab:operators}}
%	\centering
%	\begin{tabular}{rrcp{3cm}ccrc}
	\begin{tabular}{c | c | cc | c | c | c | c}
		$O_i$ & Lattice & \multicolumn{2}{c|}{Operator/Implementation} & $U=0$ & $\kappa=0$ & Error & Diagram \\ \hline
		0 	& 			& \multicolumn{2}{c|}{$\mathds{1}$} & $V$ & $V$ & 0 & \feynmandiagram[horizontal=a to b]{a -- [fermion] b}; \\ 
%			& 			& \multicolumn{2}{c|}{$V$} & & & &
%			\feynmandiagram[horizontal=a to b]{a -- [fermion] b}; \\
			&&&&&&& \\
		\vspace{-.8cm}
		1,2 	& 			& \multicolumn{2}{c|}{$p_x^\pdagger p_x^\dagger$} & $V$ & $V$ & $\bigtheta{V\delta}$ & \\
			&			& \multicolumn{2}{c|}{$\rt_x M$} & & & &
			\feynmandiagram[horizontal=a to b]{a -- [fermion, half left] b -- [insertion=0.5, half left, edge label'=$x$] a}; \\
			&&&&&&& \\
		\vspace{-.8cm}
		3,4 	& $x\sim y$ 	& \multicolumn{2}{c|}{$p_x^\pdagger p_x^\dagger h_y^\pdagger h_y^\dagger$} & $V/2$ & $V$ & $\bigtheta{V\delta^2}$ & \\
			&			& \multicolumn{2}{c|}{$\frac{2}{V}\left|\ti_x M\right|^2$} & & & & 
			\feynmandiagram[horizontal=a to b]{a -- [fermion, half left] b -- [insertion=0.5, half left, edge label'=$x$] a}; 
			\feynmandiagram[horizontal=a to b]{a -- [anti fermion, half left] b -- [insertion=0.5, half left, edge label'=$y$] a}; \\
			&&&&&&& \\
		\vspace{-.8cm}
		5,6 	& $x\nsim y$ 	& \multicolumn{2}{c|}{$p_x^\pdagger p_x^\dagger h_y^\pdagger h_y^\dagger$} & $V/2$ & $0$ & $\ordnung{V\delta^2}$ & \\
			&			& \multicolumn{2}{c|}{$\frac{2}{V}\real\ti_x M\ti_y M^\dagger$} & & & &
			\feynmandiagram[horizontal=a to b]{a -- [fermion, half left] b -- [insertion=0.5, half left, edge label'=$x$] a}; 
			\feynmandiagram[horizontal=a to b]{a -- [anti fermion, half left] b -- [insertion=0.5, half left, edge label'=$y$] a}; \\
			&&&&&&& \\
		\vspace{-.9cm}
		7,8 	& $x\sim y$ 	& \multicolumn{2}{c|}{$p_x^\dagger h_x^\dagger h_y^\pdagger p_y^\pdagger$} & $1/2$ & $V$ & $\bigtheta{\delta}$ & \\ 
		      	&           		& \multicolumn{2}{c|}{$1-\frac{2}{V}(\rt_x M+\ti_x M\mathds{P}_yM^\dagger$)} & & & &  
			\feynmandiagram[horizontal=a to b]{a [dot, label=$x\ \ $] -- [fermion, half left] b [dot, label=$\ \ y$] -- [fermion, half left] a}; \\
			&&&&&&& \\
		\vspace{-.9cm}
		9,10 & $x\nsim y$ 	& \multicolumn{2}{c|}{$p_x^\dagger h_x^\dagger h_y^\pdagger p_y^\pdagger$} & $1/2$ & $V$ & $\ordnung{\delta}$ & \\ 
			&			& \multicolumn{2}{c|}{$\frac{1}{V}(\ti_xM\mathds{P}_yM^\dagger+\ti_yM\mathds{P}_xM^\dagger$)} & & & &
			\feynmandiagram[horizontal=a to b]{a [dot, label=$x\ \ $] -- [fermion, half left] b [dot, label=$\ \ y$] -- [fermion, half left] a}; \\
			&&&&&&& \\		
		\vspace{-.8cm}		
		11,12 	& $x\sim y$ 	& \multicolumn{2}{c|}{$p_x^\pdagger p_x^\dagger p_y^\pdagger p_y^\dagger$} & $(V+1)/2$ & $V$ & $\bigtheta{V\delta^2}$ & \\
				&			& \multicolumn{2}{c|}{$\frac{2}{V}(\rt_xM+\real\left(\ti_xM\right)^2-\ti_xM\mathds{P}_xM$)} & & & &
				\feynmandiagram[horizontal=a to b]{a -- [fermion, half left] b -- [insertion=0.5, half left, edge label'=$x$] a}; 
				\feynmandiagram[horizontal=a to b]{a -- [fermion, half left] b -- [insertion=0.5, half left, edge label'=$y$] a}; \\
			&&&&&&& \\
		\vspace{-.8cm}
		13,14 	& $x\nsim y$ 	& \multicolumn{2}{c|}{$p_x^\pdagger p_x^\dagger p_y^\pdagger p_y^\dagger$} & $(V-1)/2$ & $0$ & $\ordnung{V\delta^2}$ & \\
				&			& \multicolumn{2}{c|}{$\frac{2}{V}(\real\ti_xM\ti_yM-\real\ti_yM\mathds{P}_xM$)} & & & &
				\feynmandiagram[horizontal=a to b]{a -- [fermion, half left] b -- [insertion=0.5, half left, edge label'=$x$] a}; 
				\feynmandiagram[horizontal=a to b]{a -- [fermion, half left] b -- [insertion=0.5, half left, edge label'=$y$] a}; \\
%			&&&&&&& \\
	\end{tabular}
\end{center}
\end{table*}

\begin{table*}[ht]
	\caption{Summary of physical observables (linear magnetizations and structure factors). The prime indicates that a given value can vary significantly due to finite temperature. 
	In particular $0^\prime$ only goes to zero at $\beta\rightarrow\infty$ and $V^\prime \approx \min\left(V,(\kappa\beta/\pi)^2\right)$.
	We use the Bachmann–Landau notation to denote that an error in $\bigtheta x$ scales exactly as $x$, an error in $\ordnung{x}$ scales maximally as $x$, and an error in $\littleo{x}$ is asymptotically smaller than $x$.
	\label{tab:observables}}
	\centering
	\begin{tabular}{r | c | c | c | r}
		Observable & Definition & $U=0$ & $\kappa=0$ & Error \\ \hline
		$\average{s^3_{-}}$ & $\frac 12\average{O_1-O_2}/V$ & $0$ & $0$ & $\littleo \delta$ \\
		$\average{s^3_{+}}$ & $1-\frac 12\average{O_1+O_2}/V$ & $0$ & $0$ & $\bigtheta \delta$ \\
		$S^{11}_{--}/V$ & $\frac 14\average{O_7+O_8+O_9+O_{10}}$ & $1/2$ & $V'$ & $\bigtheta \delta$ \\
		$S^{33}_{--}/V$ & $\frac 14\average{O_3+O_4-O_5-O_6+O_{11}+O_{12}-O_{13}-O_{14}}$ & $1/2$ & $V'$ & $\littleo{V\delta^2}$ \\
		$S^{11}_{++}/V$ & $\frac 14\average{O_7+O_8-O_9-O_{10}}$ & $0'$ & $0'$ & $\ordnung{\delta}$ \\
		$\average{S^{3}_{+}S^{3}_{+}}/V$ & $V-\average{O_1+O_2}+\frac 14\sum_{i\in\{3\dots 6,11\dots 14\}}\average{O_i}$ & $0'$ & $0'$ & $\bigtheta{V\delta^2}$ \\
		$S^{33}_{++}/V$ & $\average{S^{3}_{+}S^{3}_{+}}-V\average{s^3_{+}}^2$ & $0'$ & $0'$ & $\littleo{V\delta^2}$ \\
		$Q_{-}^2/V$ & $\frac14\average{-O_3-O_4+O_5+O_6+O_{11}+O_{12}-O_{13}-O_{14}}$ & $1/2$ & $0'$ & $\littleo{V\delta^2}$ \\
	\end{tabular}
\end{table*}

Let us consider the limits of no interaction ($U=0$) and no hopping ($\kappa=0$), for which the various observables can be exactly calculated for $T = 0$ ($\beta\to\infty$).
We start with the particle/hole number ($i = 1,2$ in Tab.~\ref{tab:operators}) which simply equals the number of unit cells $V$ at half-filling. Next, we consider the correlator 
between a particle and a hole ($i = 3-6$ in Tab.\ref{tab:operators}). For $U=0$, a particle does not see the holes. Therefore it is equally likely to encounter a hole 
on the same sublattice as a given particle, as it is to find a hole on the other sublattice. As there are $V$ holes in total, we find $V/2$ for each sublattice combination. 
For $\kappa = 0$, the lattice is frozen in a state of single up- or down-electrons per site with alternating sign. In terms of particles and holes, this translates to 
every $A$ site having a particle and a hole, and every $B$ site having neither (or vice-versa). Thus, a particle on a given site is perfectly correlated with all the holes on the same 
sublattice.

The coexistence of a particle and a hole as a pair in an exciton-like state ($i = 7-10$ in Tab.~\ref{tab:operators}) can be understood by a very similar consideration. 
At $U=0$ a particle is (due to charge conservation), only responsible for the existence of a single hole which can be on the same or the other sublattice yielding a correlator of $1/2$. 
$\kappa=0$ on the other hand leads to a maximal correlation between particles and holes as they can only coexist in pairs.
%Note that the existence of such a pair 
%on one sublattice implies the absence of pairs on the other sublattice and therefore comes with a maximal anti-correlation. 
%This is the reason for the sign of $S^{11}_{AB}=-O_9$.

It remains to consider correlations between particles ($i = 11-14$ in Tab.~\ref{tab:operators}) which behave completely similarly to the particle-hole correlation but for the difference 
of self-correlation. This is not relevant in the case without hopping because it is maximally correlated anyway. In the non-interacting case however there are on 
average $(V+1)/2$ particles on the sublattice that contains the given particle and $(V-1)/2$ particles on the other sublattice.

%% file: section/results.tex
%!TEX root =  ../master.tex
\section{Measurements and Extrapolation} 
By inverting the fermion matrix on stochastic Fourier sources (see~\autoref{sect:stochastic sources}) we can compute fermion propagators and use Wick's theorem to contract the propagators into the relevant observables of interest (see \autoref{tab:observables} in~\autoref{sect:15 ops}).
Our structure factors measure two-point correlations between two bilinear operators at equal time; to fully exploit each configuration we leverage time-translation invariance and average measurements on many timeslices.
Finally, the average of all measurements across the configurations within an ensemble with fixed $U$, $\beta$, Trotter discretization $N_t$, and spatial extent $L$ gives an estimator for each observable.

We find that for both the anti- and ferromagnetic spin operators the $i$=$1$ and $3$ components for $\aaverage{S^i S^i}$ give different results configuration-by-configuration and ensemble-averaged\footnote{Note that $S^{11}_{--} = S^{22}_{--}$ at the operator level.}.
This is expected since our discretization breaks chiral symmetry which would have otherwise ensured their equality \cite{Buividovich:2016tgo}.  
However, the mixed differencing of our discretization mitigates this breaking, compared to a purely-forward or purely-backward differencing.
Furthermore, in the continuum limit this symmetry is restored.
Indeed, we observe that after continuum limit extrapolation, the different components agree up to uncertainties.
We therefore directly take
\begin{equation}
m_s^2 = \frac{\sum_i S^{ii}_{--}}{V(\kappa\beta)^2}= \frac{2S^{11}_{--}+S^{33}_{--}}{V(\kappa\beta)^2},
\label{eq:definition_m_s}
\end{equation}
as the staggered magnetization (and AFMI order parameter),
simplifying the subsequent analysis.

For each $U$ and $\beta$, we perform a simultaneous continuum- and infinite-spatial-volume extrapolation.
For numerical tractability, we extrapolate extensive quantities, for example $\SAF$, which is kept finite by the IR regulation provided by the finite temperature.

%See \autoref{sect:extrap} of the supplementary material for an example of our extrapolations and a more detailed explanation of our extrapolation procedures.

\input{section/extrap}

\subsection{Data collapse}
As in Ref.~\cite{semimetalmott},  we use FSS relations to perform a simultaneous data-collapse of $m_s$ and the gap $\Delta$ and thus remove the dependence on $\beta$ in each of these quantities.
The collapse for $m_s$ is depicted in the left panel of \Figref{structure factors}.
Note that the data points below the transition, i.e.\@ $U-U_c < 0$, collapse to a single curve, whereas above the transition the points for different $\beta$ start to deviate because of the infrared cutoff imposed by our finite $\beta$ calculations.
This effect has regularly been observed before, see---for example---Fig.~7 in Ref.~\cite{White:1989zz} and Fig.~3 in Ref.~\cite{Buividovich:2018yar}.
For larger $\beta$ the collapse of the curve persists for a wider range above the transition.
Points with an obvious deviation from the global curve have been excluded from the fit so that a bias is avoided.
Since the collapse fit in $\Delta$, as done independently in Ref.~\cite{semimetalmott} yielding $U_c/\kappa=\critUold$ and $\nu=\critNuold$, requires only two free parameters $U_c$ and $\nu$, it is more stable and statistically more significant than the collapse fit of $m_s$ with the three free parameters $U_c$, $\nu$, and $\upbeta$. We therefore performed the simultaneous collapse fit of both $\Delta$ and $m_s$ giving the $\Delta$-fit a higher weight.
We find
\begin{align}
	\begin{split}
		\upbeta &= \critExp\,,\\
		\nu &= \critNu\,,\\
		\upbeta/\nu &=\critZeta\,,\\
		U_c/\kappa &= \critU\,.
	\end{split}\label{eq:final_results}
\end{align}
%For the full correlation matrix and further details see \cref{sect:covariance}.
The critical coupling is essentially unchanged from our previous result~\cite{semimetalmott}.

\input{section/covariance}

We have also considered the inclusion of sub-leading corrections to the FSS analysis, along the lines of Ref.~\cite{Otsuka:2015iba}, where
an additional factor $1+c_L L^{-\omega}$ was introduced into the scaling relation for $m_s$.
While the FSS analysis in $L$ of Ref.~\cite{Otsuka:2015iba} found a clear signal for $c_L \neq 0$, an FSS analysis
of our MC data with an additional factor $1+c_\beta \beta^{-\omega^\prime}$ turned out consistent with $c_\beta = 0$. 
For any choice of $\omega^\prime$, we found that assuming $c_\beta \neq 0$ worsened the uncertainties of the
extracted parameters, though these remained consistent with the values quoted for $c_\beta = 0$. We thus conclude that our MC data does not support the scenario of sizable 
sub-leading corrections to the scaling in $\beta$, at least within the statistical accuracy and range of temperatures covered by our present set of ensembles. 
As the (expected) Lorentz-invariance should only emerge in the vicinity of the QCP, this apparent difference in the sub-leading
corrections may reflect the underlying structure of the Hubbard Hamiltonian.

The zero-temperature AFMI order parameter $m_s(\beta=\infty)$ is also given in the right panel of \Figref{structure factors}. This has been determined from MC data extrapolated
to infinite volume and the continuum limit, using the same strategy as for the zero-temperature single-particle gap in Ref.~\cite{semimetalmott}. We have extrapolated the 
$U/\kappa=4$ values to $\beta = \infty$ and used that result to constrain the overall scale of $m_s$ (given the critical exponents we have already fixed from FSS). Note that 
$m_s(\beta=\infty)$ so determined is expected to be strictly valid only in the vicinity of the QCP.

In \Figref{s_f_and_cdw}, we show our continuum and thermodynamic limit extrapolations of the FM ($s^{ii}_{++}$) and CDW ($q^2_{-}$) order parameters.    
While $q^2_-$ is a strictly semi-positive observable, we find that several individual values of $q^2_i$ are slightly negative (though with large error bars), 
in particular for large $U/\kappa$. This effect appears to be a consequence of residual extrapolation uncertainties, as our extrapolation does not enforce semi-positivity.
Still, we observe that both $s^{ii}_{++}$ and $q^2_{-}$ clearly vanish in the zero-temperature ($\beta\to\infty$) limit for all $U/\kappa$. This behavior is in contrast to that of
$m_s$, which remains non-zero for large $U/\kappa$, and confirms the AFM character of the QCP.

%% file: section/extrap.tex
%!TEX root =  ../master.tex
\subsection{Extrapolation formulae\label{sect:extrap}}

The formulae used for the simultaneous thermodynamic and continuum limit extrapolation differ depending on the particular observable. There are however several rules applying to all of them. First of all the leading order error in the time discretisation is exactly linear $\bigtheta{\delta}$, though the coefficient is significantly reduced by our mixed differencing scheme. Next, in Ref.~\cite{semimetalmott} we showed that the spatial leading order error has to be $\bigtheta{L^{-3}}$ with the exception of observables prone to ensemble-wise positive (or negative) definite deviations. These observables cannot average out local errors over the Monte Carlo history and end up with an error in $\bigtheta{L^{-2}}$. Unfortunately the quadratic observables $S^{ii}_{--}$ and $m_s$ (\cref{eqn:extrapolation_s_af,eqn:extrapolation_m_s}) fall into this category.

From this starting point we derived the fit functions for all the magnetic observables by minor modifications. These modifications had to be made either because the leading order terms did not describe the data well enough (eq.~\eqref{eqn:extrapolation_s_1_f}), or because the in principle leading coefficient in $\delta$ (eq.~\eqref{eqn:extrapolation_s_3_s_3}) or $L$ (\cref{eqn:extrapolation_s_3_f,eqn:extrapolation_cdw}) could not be resolved numerically. A summary of the extrapolation formulae looks as follows:
\begin{align}
	\average{s^3_{+}} &= s^3_{+,0} + c^{s^3_{+}}_0\,\delta + c^{s^3_{+}}_1\,\delta \,L^{-3}\label{eqn:extrapolation_m_f}\\
	S_{AF} &= S_{AF,0} + c^{S_{AF}}_0\,\delta + c^{S_{AF}}_1\,L^{-2}\label{eqn:extrapolation_s_af}\\
	\kappa\beta m_s &= \kappa\beta m_{s,0} + c^{m_s}_0\,\delta + c^{m_s}_1 \,L^{-2}\label{eqn:extrapolation_m_s}\\
	S^{11}_{F} &= S^{11}_{F,0} + c^{S^{11}_{F}}_0\,\delta + c^{S^{11}_{F}}_1\,\delta^2 + c^{S^{11}_{F}}_2\,L^{-3} + c^{S^{11}_{F}}_3\,\delta \,L^{-3}\label{eqn:extrapolation_s_1_f}\\
	\average{S^{3}_{+}S^{3}_{+}}/V &= S^{33}_{F,0} + c^{S^{33}_{F}}_0\,\delta^2 \,L^2\label{eqn:extrapolation_s_3_s_3}\\
	S^{33}_{F} &= S^{33}_{F,0} + c^{S^{33}_{F}}_1\,\delta\label{eqn:extrapolation_s_3_f}\\
	S_\text{CDW} &= S_{\text{CDW},0} + c^{S_\text{CDW}}_0\,\delta\label{eqn:extrapolation_cdw}
\end{align}

There are two notable exceptions from the approach explained above. Firstly equation~\eqref{eqn:extrapolation_m_f} for $\average{s^3_{+}}$ can be derived analytically in the non-interacting limit. It turns out that the equation captures the leading order contributions very well even at finite coupling. $\average{s^3_{-}}$ on the other hand is always compatible with zero, even at finite $\delta$ and $L$, and is therefore not extrapolated at all.

\begin{figure}[ht]
	\includegraphics[width=0.45\columnwidth]{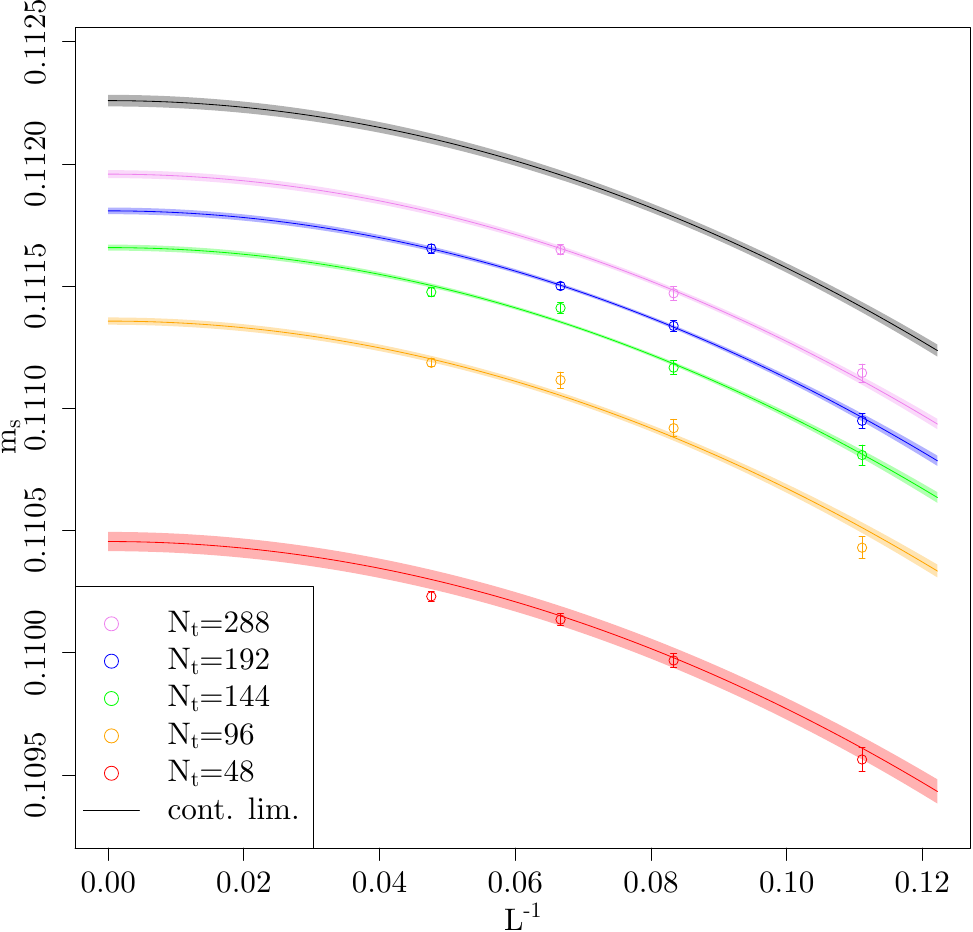}
	\includegraphics[width=0.45\columnwidth]{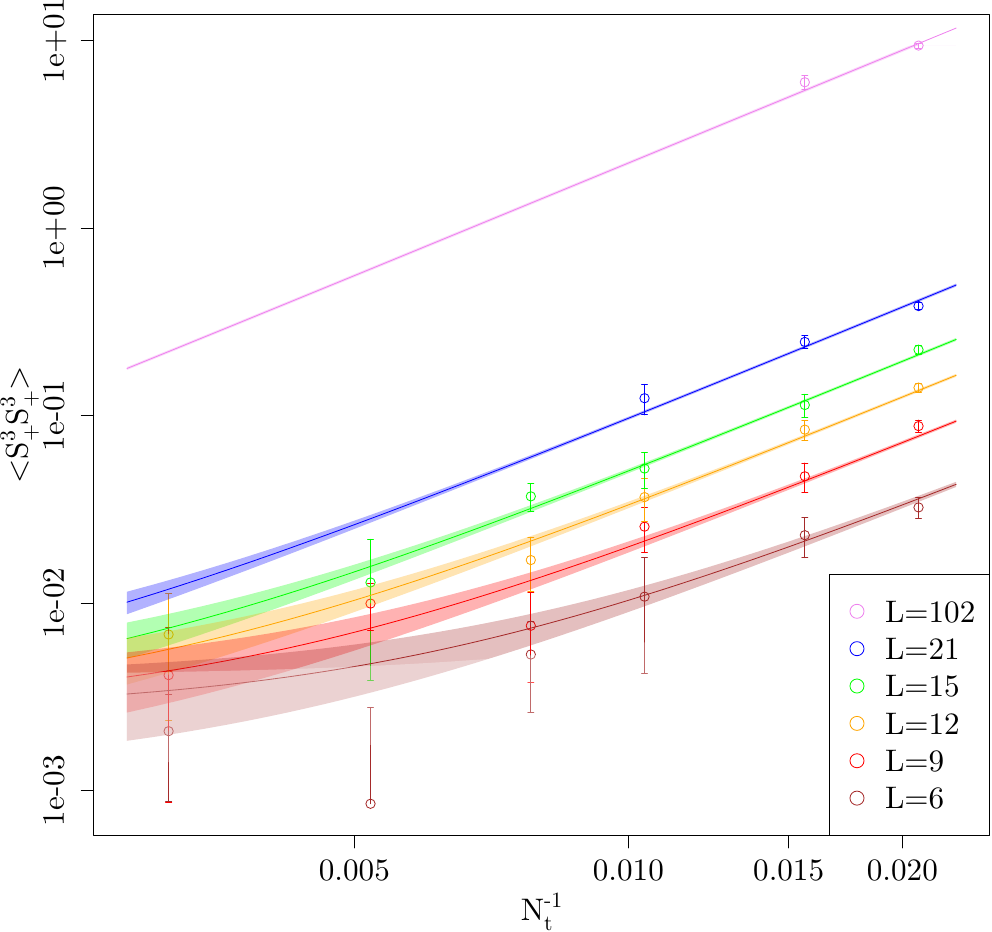}
	\caption{Simultaneous two-dimensional fit of $m_s$ using the extrapolation formula~\eqref{eqn:extrapolation_m_s} for $\beta=12$ and $U=\num{1}$ (left) and $\average{S^{3}_{+}S^{3}_{+}}$ using the extrapolation formula~\eqref{eqn:extrapolation_s_3_s_3} for $\beta=8$ and $U=\num{3.5}$ (right). All quantities are in units of $\kappa$. The fits are performed for $L\ge 9$ and $N_t/\beta\ge 6$.
	The fit on the left (right) has $\chi^2/\text{d.o.f.} \simeq \num{1.6}$ ($\chi^2/\text{d.o.f.} \simeq \num{1.1}$), corresponding to a p-value of $\simeq \num{0.08}$ ($\simeq \num{0.38}$).
	Note the double-logarithmic scale in the right panel.}
	\label{figure:extrapolation}
\end{figure}

We provide two examples for the simultaneous extrapolations in Figure~\ref{figure:extrapolation}. In the left panel we demonstrate the agreement of the data for the order parameter $m_s$ with the employed fit function~\eqref{eqn:extrapolation_m_s} even in the case of very small statistical errors.
Let us, with a look at the right panel, stress the importance of considering when to use the simple average $\average{\operator_1\operator_2}$ and when the connected version $\aaverage{\operator_1\operator_2}$. In the plot we clearly see the divergence of $\average{S^{3}_{+}S^{3}_{+}}$ in spatial volume as predicted in equation~\eqref{eqn:extrapolation_s_3_s_3}. This divergence foils any attempt of a statistically significant extrapolation, though the continuum limit would in principle yield the correct result. Should a thermodynamic limit extrapolation be attempted before the continuum limit extrapolation, one would even end up with infinite values. It is therefore of crucial importance in this case to use $S^{33}_{++}=\aaverage{S^{3}_{+}S^{3}_{+}}$ which is well behaved. On the other hand it makes sense to use $\average{\operator_1\operator_2}$ whenever $\average{\operator_1}\average{\operator_2}$ is known analytically to minimise noise. In particular $\average{S^{i}_{-}}$ and $\average{S^{1}_{+}}$ both vanish, which is why the subtraction need only be performed for $S^{33}_{++}$.

%% file: section/covariance.tex
%!TEX root =  ../master.tex
%\section{Correlation matrix of the combined collapse fit\label{sect:covariance}}

The correlation matrix
\begin{align}
	\mathrm{corr}(U_c,\, \nu,\, \upbeta) &= \matr{S[round-mode=places,round-precision = 4]S[round-mode=places,round-precision = 4]S[round-mode=places,round-precision = 4]}{
		1.00000000 & -0.03947659 & -0.1197412\\
		-0.03947659 &  1.00000000 &  0.9944137\\
		-0.11974120 &  0.99441367 &  1.0000000
	}
\end{align}
clearly shows a strong correlation between $\nu$ and $\upbeta$, whereas $U_c$ is weakly correlated. Put differently, the ratio $\upbeta/\nu=\critZeta$ can be determined with extremely high precision.
We give enough digits to yield percent-level matching to our full numerical results when inverting the correlation matrix.

%% file: section/discussion.tex
%!TEX root =  ../master.tex
\section{Discussion} 
We have presented the first fully systematic, high-precision treatment of all operators that contribute to the AFM, FM, and CDW order parameters of the Hubbard Model on a honeycomb lattice, 
followed by an FSS analysis of the SM-AFMI transition in the inverse temperature $\beta$, culminating in our results~\eqref{eq:final_results}.
Within the accuracy of our MC data, we find that the QCP associated with the SM-AFMI transition coincides with the opening of the single-particle gap $\Delta$, which disfavors
the possibility of other exotic phenomena (such as an intermediate spin-liquid phase~\cite{Meng2010,Sorella:SciRep}) 
in the vicinity of the critical coupling where the AFMI order parameter $m_s$ appears. Also, the
vanishing of the FM and CDW order parameters throughout this transition strengthens the notion that the QCP is purely antiferromagnetic in character.

In addition to an unambiguous classification of the character of the QCP of the honeycomb Hubbard model, our results demonstrate the ability to perform high-precision 
calculations of strongly correlated electronic systems using lattice stochastic methods. 
A central component of our calculations is the Hasenbusch-accelerated HMC algorithm, 
as well as other state-of-the-art techniques originally developed for lattice QCD. This has allowed us to push our calculations to system sizes which are, to date, still the largest that have 
been performed, up to 102$\times$102 unit cells (or 20,808 lattice sites).

Our progress sets the stage for future high-precision calculations of additional observables of 
the Hubbard model and its extensions, as well as other Hamiltonian theories of strongly correlated electrons~\cite{PhysRevB.100.125116,PhysRevLett.122.077602,PhysRevLett.122.077601}. 
We anticipate the continued advancement of calculations with ever increasing system sizes, through the leveraging of additional state-of-the-art techniques from lattice QCD, 
such as multigrid solvers on GPU-accelerated architectures. We are actively pursuing research along these lines.

%% file: section/acknowledgements.tex
%!TEX root =  ../master.tex

\section*{Acknowledgements}
This work was funded, in part, through financial support from the Deutsche Forschungsgemeinschaft (DFG, German Research
Foundation) through the funds provided to the Sino-German Collaborative
Research Center TRR110 “Symmetries and the Emergence of Structure in QCD”
(DFG Project-ID 196253076 - TRR 110).
E.B. is supported by the U.S. Department of Energy under Contract No. DE-FG02-93ER-40762.
The authors gratefully acknowledge the computing time granted through JARA-HPC on the supercomputer JURECA~\cite{jureca} at Forschungszentrum J\"ulich.
We also gratefully acknowledge time on DEEP~\cite{DEEP}, an experimental modular supercomputer at the J\"ulich Supercomputing Centre.
The analysis was mostly done in \texttt{R}~\cite{R_language} using \texttt{hadron}~\cite{hadron}.
We are indebted to Bartosz Kostrzewa for helping us detect a compiler bug and for lending us his expertise on HPC hardware.

%%% Local Variables:
%%% mode: latex
%%% TeX-master: "../master"
%%% End:

%% file: section/stochastic-sources.tex
\section{Stochastic Fourier Sources}\label{sect:stochastic sources}

We can estimate the trace of any matrix $D$ with dimension $d$ over a subset $I_0$ of the indices $I$ using the established method of noisy estimators~\cite{z2_noise,trace_estimation_overview} as follows. Sample a vector $\chi$ randomly with all the elements $\chi_i$ independent and identically distributed\footnote{We will later see that this is a sufficient but not a necessary condition.} (iid), such that
\begin{align}
	\erwartung{\chi}&=0\,,\\
	\erwartung{\chi_i^*\chi_j}&=\delta_{ij}\quad \text{if } i,j\in I_0\,,\label{eqn_orthonormal_noise}\\
	\chi_i&=0\quad \text{if } i\in I\setminus I_0\,.
\end{align}
Then the expectation value of $\chi^\dagger D\chi$, obtained by repeated calculation, yields the desired trace
\begin{equation}
	\tr_{I_0}D=\erwartung{\chi^\dagger D\chi}\,.
\end{equation}

\subsection{Comparison of different sources}
Ref.~\cite{trace_estimation_overview} gives an overview over the advantages and drawbacks of various distributions that can be employed for noisy trace estimation and also provides rigorous upper limits on the number of sources required to obtain a precision $\varepsilon$ with a probability of at least $1-\delta$. This upper bound in principle prefers Gaussian noise, but it is extremely loose and therefore not useful in practise. This is why we will not go into detail about these bounds. It suffices to know that such a bound exists and that the error on the trace estimation is proven to converge with $1/\sqrt n$ (as one would expect for a stochastic calculation) where $n$ is the number of noisy sources.

The approach presented in Ref.~\cite{z2_noise} gives a guideline how to identify efficient distributions that is far more useful. We present it here in a slightly modified form, but the conclusions are identical. Let us for now define $\chi^r$ to be the $r$-th noisy vector and $\erwartung{\cdot}$ denote the exact expectation value (where we usually use it for the MC-average). Then the two quantities
\begin{align}
	C_1(i) &\coloneqq \left|\erwartung{\frac1n\sum_r\left|\chi_i^r\right|^2}-1\right|\qquad \text{and}\\
	C_2(i,j) &\coloneqq \sqrt{\frac1n\erwartung{\left|\sum_r {\chi_i^r}^*\chi_j^r\right|^2}}\;,\;i\neq j
\end{align}
are good measures for the error. Here $C_1(i)$ quantifies misestimation of the diagonal entries of the matrix and $C_2(i,j)$ gives the error due to the off-diagonal elements being not suppressed perfectly.
Both $C_1(i)$ and $C_2(i,j)$ do not depend on the indices $i,j$ as long as the $\chi_i$ are identically (not necessarily independently) distributed for every $r$. We therefore drop the dependencies from now on. The expectation value of the total error for a trace estimation then reads
\begin{align}
	\sigma &= \frac{1}{\sqrt{n}}\sqrt{{C_1}^2\sum_{i\in I_0} \left|D_{ii}\right|^2+{C_2}^2\sum_{i,j\in I_0,i\neq j}\left|D_{ij}\right|^2}\,.
\end{align}
$C_1$ and $C_2$ are mostly independent from each other and we are going to minimize them individually in the following.

It is easy to see that $C_1$ can be eliminated completely by using noise on the complex unit circle, i.e.\ $\left|\chi_i^r\right|=1$ for all $i$ and $r$. This is besides other realised by (complex) Z2 noise which chooses with equal probability from $\lbrace \pm1\rbrace$ ($\lbrace\pm1,\pm\im\rbrace$).
Let us remark here that $C_1=0$ is a very important feature because noisy trace estimation is only efficient for diagonally dominated matrices. Z2 and similar noise calculate the trace of a diagonal matrix exactly with a single source whereas other distributions, e.g.\ Gaussian noise, would still require a high number of sources to get a decent approximation.

On the other hand $C_2$ is nearly independent of the distribution. In the case where all $\chi_i^r$ are iid, one finds
\begin{align}
	\erwartung{\eta^r} &= 0\,,\\
	\erwartung{\left|\eta^r\right|^2} &= \erwartung{\left({\chi_i^r}^*\chi_j^r\right)^*{\chi_i^r}^*\chi_j^r}
	= \erwartung{\chi_i^r{\chi_i^r}^*}\erwartung{{\chi_j^r}^*\chi_j^r}
	= 1
\end{align}
where we defined $\eta^r\coloneqq{\chi_i^r}^*\chi_j^r$ and dropped the indices $i\neq j$ as before. Thus $\eta^r$ is iid with zero expectation value and unit standard deviation. The central limit theorem now guarantees that for large $r$ the distribution approaches $\sum_r\eta^r \sim \mathcal{N}_{0,n}$. Moreover the variance is an additive quantity for independent variables which yields $C_2=1$ exactly.

This leads to the conclusion that any iid noise with numbers of unit modulus gives the same error and this error is the theoretically optimal one for iid noise.

In practice one is often not interested in the complete trace but for example in its real part only. Then real iid sources give the well~known result $C_2=1$. Remarkably however the same result also holds for complex noise where the deviation is split into the real and imaginary parts $C_2^R$ and $C_2^I$ respectively as ${C_2}^2={C_2^R}^2+{C_2^I}^2$. We observe that both projections can be minimised when $C_2^R=C_2^I=1/\sqrt{2}$. This is the case in complex Z2 noise which is the reason why it is usually preferred over real Z2 noise.

Note that neither of the possible error contributions $C_1$ and $C_2$ contains the condition $\erwartung{\chi}=0$ nor independence or identical distribution ($\erwartung{\eta^r}=0$ in contrast is required). We are therefore going to drop these superfluous requirements and see if we can find an even more efficient method for trace estimations. As $C_1$ cannot be optimised any further, we only have to consider $C_2$ which is the average of all variances and covariances of $\eta^r$ over $r$. We are now only considering noise with $\left|\chi_i^r\right|=1$, so $\left|\eta^r\right|=1$ as well and the variance of $\eta^r$ is guaranteed to be 1 as before. Therefore the only variable one can optimise is the covariance of $\eta^r$
\begin{align}
	\zeta^{rs} \coloneqq \mathrm{cov}\left({\eta^r}^*,\eta^s\right)
\end{align}
with which we can rewrite the off-diagonal error contribution
\begin{align}
	C_2 &= \sqrt{\frac1n\erwartung{\sum_{r,s}{\eta^r}^*\eta^s}}
	= \sqrt{\frac1n\erwartung{\sum_r\left|\eta^r\right|^2 + \sum_{r\neq s}{\eta^r}^*\eta^s}}
	= \sqrt{1+\frac1n\sum_{r\neq s}\zeta^{rs}}\,.
\end{align}
The symmetry in $r$ and $s$ suggests that $C_2$ is minimised when all $\zeta^{rs}$ are equal and maximally negative. This optimum is, quite intuitively, reached for all $\chi^r$ pairwise orthogonal. In this case the probability $p_r$ of redundancy one would have for iid noise is reduced to $p_r-1/(d-1)$ because one of the dimensions cannot be reached again. This difference already is the covariance we are looking for. Thus we obtain a lower bound for the error contribution
\begin{align}
	C_2 &\ge \sqrt{1+\frac1n\sum_{r\neq s}\frac{-1}{d-1}}
	= \sqrt{1-\frac{n(n-1)}{n(d-1)}}
	= \sqrt{1-\frac{n-1}{d-1}}
	= 1-\frac{n}{2d}+\ordnung{d^{-1}}+\ordnung{\left(\frac nd\right)^2}\,.
\end{align}
It is easy to convince oneself that this really is the optimum because the error vanishes for $n=d$ when one has sampled the full space. Therefore there is no deterministic nor stochastic algorithm that does not use any specific features of a given matrix and is more efficient than this method.

An efficient practical realisation would be to draw
\begin{align}
	\chi_i^0 &\sim \left\lbrace\eto{\im2\pi\frac kd} | k\in\lbrace0,\dots,d-1\rbrace\right\rbrace
\end{align}
iid and for all $r\ge1$ to sample $j_r\in\lbrace1,\dots,d-1\rbrace$ without repetition setting
\begin{align}
	\chi_i^r &= \eto{\im2\pi\frac {i}{d}j_r}\chi_i^0\,.
\end{align}

Last but not least we emphasise that for the trace estimation to be useful one has to reach the desired accuracy at some $n\ll d$. So the tiny gain from choosing all noise vectors pairwise orthogonal is negligible in practice.

\subsection{Second order observables}
The quartic operators $O_{3\dots14}$ cannot be expressed in terms of simple traces. Instead we have to generalise the noisy source estimation.
There are different possible approaches. In our simulations we chose to sample $\chi$ again as above, then define
\begin{align}
	\chi' &= M^{-1}\chi\,,\\
	\chi''&={M^{\dagger}}^{-1}\chi
\end{align}
and estimate the relevant quantities as
\begin{align}
	\sum_{x,y\in A}\left|\left(M^{-1}\right)_{(x,t),(y,t)}\right|^2 &= \erwartung{\sum_{x\in A}\left|\chi'_{x,t}\right|^2}\,,\\
	\sum_{x,y\in A}\left(\left(M^{-1}\right)_{(y,t),(x,t)}\left(M^{-1}\right)_{(x,t),(y,t)}\right) &= \erwartung{\sum_{x\in A}{\chi''_{x,t}}^*\chi'_{x,t}}\,.
\end{align}
Note that the right hand side term is not a simple scalar product $\chi'^\dagger\cdot \chi'$, respectively $\chi''^\dagger\cdot\chi'$. We have to include the projection here explicitly. This allows to compute $\sum_{x\in A,y\in B}$ nearly without additional computational effort, as only the projection sum has to be adjusted, not the linear solve.

\subsection{Average over time slices}
For the structure factors the averages over the time slices have to be included explicitly and cannot be absorbed in the traces. This means that the number of linear solves increases from the number of noisy sources $n$ to $N_t\cdot n$. As this is extremely expensive, we evaluate the average over all time slices by an average including only a fixed number of equidistant times, which is guaranteed to be equal in the infinite-statistics limit by translation invariance while saving computational cost without losing much accuracy, since neighbouring time slices on any given configuration are correlated.

Another approach would be to sample $\chi$ not from e.g.\@ $I_0=(A,t)$, but from $I$ always and include the projection later
\begin{equation}
	\tr_{I_0}D=\sum_{i\in I_0,j\in I}\erwartung{\chi_i^*D_{ij}\chi_j}\,.\label{eqn_trace_projected}
\end{equation}
Na\"ively one would expect that this saves a lot of linear solves, but it also introduces a lot of additional noise. The variance of the sum in equation~\eqref{eqn_trace_projected} increases linearly in the cardinality of $I$. In order to compensate this increase in statistical fluctuations we would have to increase the number of noisy sources by the same amount. Thus at equal statistical precision the number of linear solves stays the same.

\subsection{The time-shifted trace}
In general we do not only have to calculate traces, but also sums over off-diagonal matrix elements---correlators between operators at different times. This time shift can be implemented in the following way (again $\chi$ sampled as above):
\begin{align}
	\sum_t D_{t+\tau,t}&=\sum_{t,t'}D_{t',t}\delta_{t',t+\tau}
	=\sum_{t,t'}D_{t',t}\erwartung{\chi_{t'}^*\chi^{\vphantom{*}}_{t+\tau}}
	=\erwartung{\sum_t \left(D^\dagger\chi\right)^*_t\chi^{\vphantom{*}}_{t+\tau}}\,,
\end{align}
where $\tau$ is an arbitrary time shift.